
\magnification 1200
\def\blankline{\par\vskip 12 pt\noindent}
\def\ref{\par\noindent\hangindent 20pt}
\parindent 15pt
{}~
\vskip 2truecm
\centerline{\bf  BRIGHT STARS IN THE GALACTIC GLOBULAR M5}
\vskip 2truecm
\centerline{\bf E.Brocato$^1$, V.Castellani$^1$ $^2$, V.Ripepi$^2$}
\vskip 1truecm
\centerline{$^1$Osservatorio Astronomico Collurania, I-64100 Teramo, Italy}
\centerline{$^2$Dipartimento di Fisica, University of Pisa, I-56100 Pisa,
Italy}
\vskip 3truecm
\centerline{Version 5: June 1994}
\bigskip
\bigskip
\bigskip
\bigskip
\centerline{Submitted to Astronomical Journal}
\vskip 3truecm
\noindent
\blankline
\blankline
\blankline
\vskip 2truecm
{\it This paper is based on observations obtained at the
European Southern Observatory, La Silla, Chile}
\eject
\baselineskip=24pt
\noindent
{\bf  Summary}
\par
We present a CCD investigation of the
galactic globular M5 aimed to increase the statistical
relevance of the available sample of evolving bright stars.
Previous investigations, limited to the outer cluster region,
have been extended toward the cluster center, more than
doubling the number of observed luminous stars.
On this basis, we discuss a statistically relevant sample,
rich of 415 HB stars. The occurrence
of a gap in the blue side of the HB is suggested.
Comparison to
the current evolutionary scenario discloses a good agreement
concerning both the C-M diagram location and the relative
abundance of stars in the advanced evolutionary phases,
supporting our present knowledge
of the evolution of low mass stars.
Determination of the amount of the original helium content
through the ratio R N(HB)/N(RGB) gives
$Y = 0.22 \pm 0.02$.

\par

\blankline

{\bf 1. Introduction}
\par
As well known, current evolutionary theories concerning post main
sequence evolution of globular cluster stars give relevant predictions
not only about the CM location of the stars but also about
their number distribution
over the advanced  evolutionary phases.
The abundant literature on that matter already disclosed a
general agreement between such a distribution and theory, supporting
the current evolutionary scenario. However, this kind of
investigation finds its natural limit in the statistical fluctuations
affecting observational samples,
so that any increase in the number of the observed stars
implies an increase in the accuracy of the
results, producing more stringent constraints to the theoretical
frame.
\par
Owing to the relatively large portion of sky covered by a galactic
globular cluster, statistically significant investigations of cluster
 giants
has been often  produced on the basis of photographic material,
small field CCD techniques having mainly devoted to deep photometric
investigations of selected portions of a cluster. According to such
an occurrence, one finds that available investigations
on the population of bright stars in globular clusters
are typically limited to the outer regions of the
clusters, neglecting the central regions where the crowding
of the stellar images do not allow accurate photographic photometry.
\par
However, CCD photometry joined to the modern procedures
for image analysis allows a much deeper penetration in the
crowded central fields, offering  an ideal tool to complement
the previous samples with rich populations of bright stars.
This paper reports the result of a similar investigation,
devoted to study luminous stars in the well known cluster
M5 (=NGC 5904). The choice of the target and the production
of the observational material were actually
driven by a parallel investigation devoted to produce
accurate curve of light for RR Lyrae variables in this cluster.
However, since cluster luminous
stars have been throghly studied by Buonanno  et al.(1981)
all over the region $r~~>~~ 120~~arcsec$,
 we took advantage of our CCD coverage
of the central region to improve the C-M diagram with new data,
significatively enlarging the  available sample of objects.

\par
As we will discuss in the following sections, our CCD material
allow to extend the investigation down to 20 arcsec from the cluster center,
more than doubling the number of observed cluster giants. Next
section will describe the observation and the reduction procedure.
Sections 3 and 4 will deal with the discussion of the results, comparing
observational data with  evolutionary prescriptions for both
H shell burning and He burning phases. We will find that both
the number distribution of stars and their HR diagram location
appear in agreement with theoretical prescriptions.
On this basis, in the final section we will present
an updated evaluation of the cluster original He content.

\blankline

{\bf 2. The observations.}
\par
Observations have been performed at the 1.5 m Danish ESO telescope
at La Silla during the night from 14 to 17 April 1989. The chip
was RCA 512x331 , 0.5 arcsec/pixel sized. The cluster region was
covered with four overlapping fields, as shown in fig. 1. Instrumental
magnitudes were obtained on the basis of 4 exposures in
each filter and for each field.

Exposure times were 40 sec for V filter
and 90 sec for B. Flat fields and bias exposures were obtained at the
beginning and at the end of each night to correct the CCD response
to uniform sensitivity.
\par
Data reduction was performed using ROMAFOT package for crowded
photometry. All the magnitudes in the frames for a given field and
with a given filter were normalized to the instrumental magnitudes
of a reference frame, taking their average as the final value.
These instrumental magnitudes were finally calibrated  by comparison
with 88 stars in common with the CCD investigation by Storm, Carney
\& Beck (1991) with $V < 17$. As a result we found
\par
{}~~~~~~$V= v + 27.13 \pm 0.02$.

{}~~~~~~$B-V= (1.097 \pm 0.009) (b-v) + 0.42 \pm 0.02$.
\par
with an internal error smaller than $0.05 mag$ in both colors for
$V \le 18~~ mag$.
\par
Completness was tested by randomly adding to the original frames a set
of stars of various magnitudes and colors. In this way we found that the
sample of luminous stars ( $V \le 17~~ mag$) was fairly complete
down to 20 arcsec from the cluster center. Thus in the following we will
refer only to the  sample of 3736 stars located beyond the quoted limit.
Data are directly available by E-mail request to the Authors or
via Xmosaic at Teramo observatory WWW server.
Fig. 2 shows the C-M diagrams of stars in selected anuli and the
corresponding diagram for the whole sample of stars.
\par
A quick look to the last figure shows three interesting features:

{\it1)} the well defined location of the AGB clump;
{\it 2)} the evidence for the RG bump and, finally,
{\it 3)} the probable but theoretically unespected occurrence of a gap
in the HB distribution at $B-V \simeq 0$ and $V \simeq 15 $.

If this last point is real, thus M5 has to be added to the cluster
(like M 15, see Buonanno et al 1981) showing this unexplained feature.

\par

\blankline

{\bf 3. Theoretical scenarios}
\par
According to the current evolutionary scenario, the abundance
of stars in a given  advanced evolutionary phase is
governed by the time spent by stars in that phase, resulting directly
proportional to this time. On this basis, one finds in the literature
stringent theoretical expectations  concerning the distribution of stars
along the Red Giant (RG) branch of a globular (see, e.g.,
Ferraro et al 1992 and references therein). According to the same
philosophy, since the pioneering paper by Iben (1968) we know
that the number ratio (R) of HB stars to RGs
at luminosities larger than the HB luminosity level gives information
about the amount of original Helium in cluster stars. We can  add
that detailed computations of post HB evolution, as given in
Castellani, Chieffi \& Pulone 1991 (herein after CCP),
 already produced theoretical
expectations also about the number ratio between HB and Asymptotic
Giant Branch (AGB) stars . Thus the comparison of all these expectations
with a rich sample of luminous stars will be of great interest.
\par
However, before approaching such a comparison it is worth to
recall  in some detail the theoretical frame. One  may notice that
the observational parameters  one is dealing with
do  have an internal connection.
As a matter of fact, the evaluation of the amount of original He
through the R-method implies the following three assumptions:

i) Theoretical estimates of the evolutionary times along the RG
branch are correct,

ii) Similar extimates for the evolutionary times in the HB phase are
also correct.

iii) The luminosity of  HBs do follow the canonical dependence on
the amount of original He.

As for the first point,  RG lifetimes appear a well
established theoretical result (see, e.g., the discussion in
Castellani \& Norci 1989, hereinafter CN) which only needs comparison
with observations. The next two points require more attention.
\par
Current canonical theories
concerning HB evolution and ,thus, HB lifetimes assume
the efficiency of semiconvection with a negligible efficiency of
the so called "breathing pulses". However, it
has been suggested that semiconvection could be overcome by the
efficiency of a strong overshooting (Bressan, Bertelli \& Chiosi 1989).
In general, one expects that any variation in the estimates of convective
mixing in the stellar core
could deeply affect HB lifetimes. Comparison with RG lifetimes
(Caputo et al 1989) already supported the reliabilty of the
canonical semiconvective scenario. The availability of
canonical prescriptions about post HB  evolutionary
phases allow now a further relevant test on the matter.
A more efficient mixing increases the mass of the He depleted core at the
end of HB evolution and a larger HB lifetime is consequently obtained.
Moreover, the luminosity of the bottom level of the AGB is also increased
and
the number ratio between AGB and HB stars decreases, with respect to
the value fixed by canonical
evaluation at $N_{AGB}/N_{HB} = 0.12 \pm 0.01$ (CCP).
\par
The actual luminosity of HBs has been finally seriously debated
in the current literature. Evidences for the so called Sandage's
effect (see Sandage 1993 and reference therein) and/or the results
of Baade Wesselink analyses of RR Lyrae stars raised the suggestion
for a discrepancy between canonical predictions and the
actual luminosity of these HB stars. Even if, in our feeling, there
is an increasing evidence against a similar occurrence (see, e.g.,
Castellani, Degl'Innocenti \& Luridiana 1993: CDL) one has to bear in mind
all this scenario to approach a correct discussion of  to the data we
will deal with.
\par
\blankline
{\bf The Red Giant Branch}
\par
To follow the various arguments in the quoted order, let us discuss first
RG branch stars. If N represents the
number of giants in a given interval of luminosity, thus theoretical
predictions can be put in the form
\par
{}~~~~~~~~$log~ N = A - B ~log~ L + C (L_b$)
\par
where A is a constant of normalization to the total number of observed
stars whereas the slope B is predicted by the theory. $C(L_b$) is an
additional term representing the "bump"  superimposed
to the general distribution at the luminosity $L_b$ where
the H burning shell encounters the H discontinuity in the
stellar interior. In principle,
comparison of the slopes does not require a precise determination
of the absolute magnitudes Mv and, in turn, of the cluster distance
modulus DM. However, to estimate the luminosity  of the bump one needs
such an evaluation.
To this purposes, let us adopt the distance modulus
derived assuming for the HB the canonical luminosity
as derived for the suitable choice about the
 chemical composition Z=0.001, Y=0.23.
On this basis one can estimate for M5 $DM = 14.4 \pm 0.1$
(CCP).
\par
The large sample of giants we are dealing with allows
the analysis of the distribution
of the giants binned in $0.1 ~~mag$ intervals, which
 appears the most appropriate
procedure to compare observation with theoretical constraints,
as more deeply discussed in CN.
Fig.3 compares the observed distribution with the
corresponding theoretical prediction as obtained from a synthetic RG branch
populated with the same number of stars as observed.
Inspection of this figure reveals  a good correspondence between observed
and predicted slope of the distribution, an occurrence which supports
theoretical evaluations concerning the core mass versus luminosity relation
for RG structures and, as a consequence ,
provides a strong support to
the general reliability of theoretical evolutionary times for low
mass red giants.
\par
As a further relevant feature, one finds that the RG "bump" is
clearly detected and located in the interval V 14.9 - 15.0 m, about
0.4 magnitudes fainter than what expected by theoretical side. This is
a well known occurrence common to other galactic globulars, widely
discussed by Fusi Pecci et al (1990).
On evolutionary grounds, this can be
taken as an evidence that surface convection sinks down a bit more
than expected in canonical computations, with negligeable
outcome on the general evolutionary scenario. Here let us only remind that
such a discrepancy can be, at least in part, connected to a
not solar ratio of heavy elements in Population II stars,
with enhancement of the alpha-elements.( e.g. Salaris et al
1993) \par
However, comparison with
theoretical prescriptions, as given in CN, discloses a series of
interesting concordances. As a first point, one finds that the
distribution in luminosity agrees with theoretical predictions, for which the
bump should not exceed the typical width  of about 0.1 magnitudes.
Even more interesting, one finds that the amplitude of the bump
closely follows the predictions given in CN. According to the
analysis given in that paper,  for the assumed
cluster chemical composition (Y=0.23, Z=0.001) and for a cluster age
of the order of 15 billion years, one expects a "contrast" of
the bump above the continuous sloping distribution as given
by $D ~logN \simeq 0.5$, as observed.
\par
In this respect, we can conclude that the distribution of stars along
the red giant branch of M5 gives a satisfactory
support to current evaluations about red giant evolution, which
can  be regarded among the best established evolutionary
phases of low mass stars. The discrepancy in the bump luminosity remains
 an open question
which deserves further investigation. Here we wish only to notice that
the observed bump, though fainter than expected, remains
above the luminosity of the horizontal branch, still contributing to the
number of giants to be taken into account in the evaluation of the
parameter R. Accordingly, the fainter magnitude should have
a negligeable influence on the total number of giants, and we will
hold on the R calibration given by Buzzoni et al (1983) on the basis
of canonical computations.

\par
\blankline
{\bf 4. The He burning phases.}
\par
As a first evolutionary test,
let us approach the problem of the HR diagram location of
He burning stars by comparing in Fig.4 the observed C-M diagram
for luminous stars with the corresponding theoretical predictions as
taken from CCP. Buonanno  et al. (1981) discussed possible evidences for
a disagreement between theory and observations, as given by the
suggested overluminosity of the bluest portion of the  observed HB branch.
However, fig.4 shows that theoretical evolutionary tracks, when translated
into the color magnitude diagram by adopting color temperature relations
and bolometric corrections from Kurucz (1979), fairly fit the observed
distribution all along the horizontal branch, thus overcoming the
quoted problem.
\par
As a very important point, let us notice that  fig. 4 shows a
close agreement not only for the HB phases but also, and in particular, for
the luminosity of the clump of AGB stars marking the initial
phases of double shell He burning phases. As already discussed,
this is an evolutionary parameter which critically depends on the
extension of central convective mixing in the previous central
He burning structures. The observed agreement gives thus
a relevant  support to the adopted treatment of semiconvection in He
burning stars.
\par
As discussed in a previous section, the number ratio between
HB and AGB stars is another indicator of the efficiency of
convective or semiconvective mixing during the major phase of central He
burning. Canonical evaluations performed under the
assumptions $Y = 0.23$  (CCP) give the prediction $N_{AGB}/N_{HB} = 0.12 \pm
0.01$, independently of the assumptions about the cluster metallicity or
about the color distribution of HB stars. To discuss these and other
features of HB stars in M5 we report in Table 1 selected quantities
derived from our observational results together with similar results
as presented by Buonanno  et al. (1981) for stars in the external
region of the cluster. As for our results, table 1 gives separately
data for $r \leq 120~~ arcsec$, to be added to Buonnanno's
sample to produce the "complete" sample, and data for $r \geq 120
{}~~arcsec$  wich extend our sample to outer regions already covered
by Buonanno et al.
\par
It appears that the present investigation is more than doubling
the number of He burning stars found in the cluster, with
a significant improvement on the statistical relevance of the sample.
As for the number ratio of AGB to HB stars, our sample gives
$N_{AGB}/N_{HB} = 0.119 \pm 0.029$,
a figure which closely overlaps theoretical predictions.
However, if one complements the data by Buonanno et al. with our data
for $r \leq 120 ~~arcsec$, one finds a sligthly different result, namely
$N_{AGB}/N_{HB} = 0.147 \pm 0.026$. As a matter of fact, data in table 1
disclose that the samples we are now dealing with, have a similar number
of AGB stars, our sample being  richer of HB stars than Buonanno et al.
This is a rather curious occurrence, since our samples for $r \le 120
{}~~arcsec$ and for $r \ge 120$ have very similar values for this ratio.
Thus it appears that the overabundance of AGB stars in Buonanno's
sample is just coming from the region of cluster outside our
investigation.
\par
Hovever, one finds that the number ratio between AGB and HB
stars in the complete sample still remains  in agreement
with theoretical predictions within 1 sigma. Since any discussion
within the quoted error is not really meaningful, one can only conclude
that theory and observation appear in agreement also for such a relevant
evolutionary parameter.
One may conclude that evolving stars in the galactic globular M5
appears to be distributed in the HR diagram according to all the
available theoretical prescriptions, an agreement which
allow us to be confident in our theoretical understanding
the evolution of low mass stars in galactic globulars.
\par
On this basis, one becomes also more and more confident in the
theoretical scenario supporting the calibration
given by Buzzoni et al (1983) of the parameter R as a function of the
original He content in cluster stars. According to the normal procedure,
let us define R as the number ratio between HB stars and RG stars
more luminous than the mean bolometric luminosity of the
HB at the RR Lyrae gap. Adopting $V = 15.10 ~~mag$ as the appropriate HB
luminosity level and  $V = 15.25 ~~mag$ as the corresponding $V$ magnitudes
along the RG branch, from the ratio of the corresponding counts as given in
Table 1 one derives

{}~~~~~~~~~~~~~~~~$R = 1.28 \pm 0.13$

where the error (1 $\sigma$) account for
the expected statistical fluctuations of the
counts. From our data, one finds that a variation of
$\pm 0.1 ~~mag$ in the lower luminosity boundary  of the RG sample gives a
variation of $\pm 0.03$ in R. Thus the statistics is still the main
source of indetermination in this analysis. By adopting the Buzzoni
et al (1983) calibration, one finally finds

{}~~~~~~~~~~~~~~~~$Y= 0.22 \pm 0.02$

which we present as an updated evaluation of the
original amount of He in stars
membering the galactic globular M5.
\par
\blankline
{\bf 5.Conclusions}
\par
In this paper we presented a new CCD investigation of luminous stars in the
galactic globular cluster M5, extending to the cluster central regions
a previous investigation given by Buonanno  et al.  (1981) on
post main sequence evolutionary phases.  On this basis, we
secured a statistically relevant sample, rich of 415 HB stars,
to be compared  to theoretical constraints concerning the number
of stars populating RG, HB and AGB branches.
\par
The comparison revealed a general agreement between theoretical
prediction and the observed numbers of stars, supporting the
results of current theoretical  computations concerning the
evolution of low mass stars both during the H and the He burning phases.
A further support to these computations has been also found
in the good theoretical fitting of the HB CM diagram location
and, in particular, in the correct prediction of the luminosity
of the AGB clump.
\par
According to such an agreement, we present a new evaluation of
the original He content of cluster stars, as given by the
value $Y ~=~0.22 ~\pm~ 0.02$ which reconciles previous evaluation
presented by Buonanno et al. (Y=0.17) with the general
belief of an amount of cosmological He of the order of Y=0.23.

\vfill

\eject

\nopagenumbers
\noindent

{\bf References}
\blankline

\ref {Bressan A., Bertelli G. \& Chiosi C. 1986, Mem. Soc. Astron. It. 57,411}
\ref {Buonanno R., Corsi C.E. \& Fusi Pecci F. 1981, MNRAS, 196,435}
\ref {Buzzoni A., Fusi Pecci F., Buonanno R. \& Corsi C.E. 1983, A\&A 128, 94}
\ref {Caputo F., Castellani V.,Chieffi A. \& Pulone L. 1989, ApJ 340,241}
\ref {Castellani V., Chieffi A. \& Pulone L. 1991, ApJ Suppl. , 76, 911, CCP}
\ref {Castellani V., Degl'Innocenti S. \& Luridiana V. 1993, A\&A 272, 442)
\ref {Castellani V. \& Norci L. 1989, A\&A 216, 62}
\ref {Ferraro F.R., Fusi Pecci F., Buonanno R. 1992, MNRAS 256,376}
\ref {Fusi Pecci F., Ferraro F.R., Crocker D.A., Rood
R.T., Buonanno R. 1992, A\&A 238,95}
\ref {Iben,I.Jr. 1968, Nature 220,143}
\ref {Kurucz 1979 ApJ Suppl. 49,1}
\ref {Sandage A. 1993, A.J. 106, 687}
\ref {Storm J., Carney B.W. \& Beck J.A. 1991, PASP 103,1264}
\ref {Salaris M., Straniero O. \& Chieffi A. 1993, ApJ 414,580}
\vfill
\eject

\centerline{\bf Figure captions}
\blankline \blankline
\ref{Fig. 1:  The sky location covered by our four CCD frames in the field
of the cluster M5. The circle (r=120 arcsec) shows the inner boundary
of the region photographically explored by Buonanno  et al. (1981)}
\ref {Fig.2: The C-M diagram for stars in selected regions of
the cluster, as labeled, and for the total sample of measured stars.
Arrows indicate the three features discussed in the text.}
\ref {Fig.3: The observed distribution with luminosity of the number
of red giants binned in 0.1 mag intervals (full line). Dashed line shows the
result of a theoretical simulation as obtained populating a red giant
branch with the same number of stars as observed.}
\ref {Fig.4: The observed C-M diagram for our sample of luminous stars in M5
compared with theoretical evolutionary tracks from CCP for Z=0.001, Y=0.23}
\vfill

\eject

\par
\par
Table 1: Number of cluster giants in the various evolutionary phases and for
selected cluster regions, as labelled.

\par
\par

\par
{}~N(HB)~~~~N(RGB)~~~~N(AGB).

{}~251~~~~~~~184~~~~~~~~30~~~a) This paper, $20 < r < 120 arcsec$.

{}~164~~~~~~~140~~~~~~~~31~~~b) Buonanno et al.,$ r > 120 arcsec$.

{}~415~~~~~~~324~~~~~~~~61~~~c) Complete sample, a + b.

{}~~77~~~~~~~~84~~~~~~~~12~~~d) This paper,$ r > 120 arcsec$
\vfill

\bye